\documentclass[prb,amsmath,superscriptaddress,showpacs,showkeys,twocolumn]{revtex4-1}
\usepackage{amssymb,amsmath}   
\usepackage[dvips]{graphicx}   
\usepackage{verbatim}   
\usepackage{color}      
\usepackage{subfigure}  
\usepackage{hyperref}   
\usepackage{gensymb}
\usepackage{epstopdf}
\usepackage{natbib}
\usepackage{enumerate}
\usepackage{mathbbol}
\usepackage{braket}
\DeclareMathOperator{\Tr}{Tr}

\begin{document}
\title{Spin-dependent transport through a chiral molecule in the presence of spin-orbit interaction and non-unitary effects}
\author{Shlomi Matityahu} \email{matityas@post.bgu.ac.il}
\affiliation{Department of Physics, Ben-Gurion University, Beer
Sheva 84105, Israel}
\author{Yasuhiro Utsumi}
\affiliation{Department of Physics Engineering, Faculty of
Engineering, Mie University, Tsu, Mie, 514-8507, Japan}
\author{Amnon Aharony}
\affiliation{Department of Physics, Ben-Gurion University, Beer
Sheva 84105, Israel} \affiliation{Ilse Katz Center for Meso- and
Nano-Scale Science and Technology, Ben-Gurion University, Beer
Sheva 84105, Israel} \affiliation{Raymond and Beverly Sackler
School of Physics and Astronomy, Tel Aviv University, Tel Aviv
69978, Israel}
\author{Ora Entin-Wohlman}
\affiliation{Department of Physics, Ben-Gurion University, Beer
Sheva 84105, Israel} \affiliation{Ilse Katz Center for Meso- and
Nano-Scale Science and Technology, Ben-Gurion University, Beer
Sheva 84105, Israel} \affiliation{Raymond and Beverly Sackler
School of Physics and Astronomy, Tel Aviv University, Tel Aviv
69978, Israel}
\author{Carlos A. Balseiro}
\affiliation{Centro At\'{o}mico Bariloche and Instituto Balseiro,
Comisi\'{o}n Nacional de Energ\'{i}a At\'{o}mica, 8400 S. C. de
Bariloche, Argentina} \affiliation{Consejo Nacional de
Investigaciones Cient\'{i}ficas y T\'{e}cnicas (CONICET),
Argentina}
\date{\today}
\begin{abstract}
Recent experiments have demonstrated the efficacy of chiral helically shaped molecules in polarizing the scattered electron spin, an effect termed as chiral-induced spin selectivity (CISS). Here we solve a simple tight-binding model for electron transport through a single helical molecule, with spin-orbit interactions on the bonds along the helix. Quantum interference is introduced via additional electron hopping between neighboring sites in the direction of the helix axis. When the helix is connected to two one-dimensional single-mode leads, time-reversal symmetry prevents spin polarization of the outgoing electrons. One possible way to retrieve such a polarization is to allow leakage of electrons from
the helix to the environment, via additional outgoing leads. Technically, the leakage generates complex site self-energies, which break unitarity. As a result, the electron waves in the helix become evanescent, with different decay lengths for different spin polarizations, yielding a net spin polarization of the outgoing electrons, which increases with the length of the helix (as observed experimentally). A maximal polarization can be measured at a finite angle away from the helix axis.
\end{abstract}

 \maketitle
\section{Introduction} \label{Sec1}
One of the most promising subfields of spintronics is organic spintronics,\cite{NWJM07,DVA,BC13} which exploits organic materials to manipulate and control spin currents. Recently, this emerging field has experienced a significant progress with the remarkable discovery of spin-dependent transport through chiral organic molecules at room temperature. This so-called chiral-induced spin selectivity (CISS) effect has been first observed in double-stranded DNA molecules (dsDNA).\cite{RSG06,GB11,XZ11} Whereas dsDNA molecules were found to act as highly efficient spin filters, no CISS effect was found in single-stranded DNA (ssDNA) molecules.\cite{GB11} However, recent experiments found spin polarization of electrons
transmitted through a bacteriorhodopsin (bR),\cite{MD13,EH15} a protein composed of seven parallel single $\alpha$-helices, when the bR was embedded in a purple membrane which was adsorbed on a variety of substrates. These experiments have proved that the CISS effect results from the intrinsic properties of the helical molecules, and is almost independent of the substrate.\cite{MD13} Additionally, a self-assembled monolayer of a different $\alpha$-helical protein was used to demonstrate the operation of a chiral-based magnetic memory.\cite{BDO13}

Following these experimental observations, several groups have attempted to explain the theory behind the CISS effect.\cite{GR12,GR13,GAM12,GAM14,VS14,ME12,EAA13,GJ13} In one approach, tight-binding models are used to study electron transport through helical molecules in the presence of spin-orbit interaction (SOI).\cite{GR12,GR13,GAM12,GAM14} In a second
approach, a spin-dependent scattering problem is solved in a three-dimensional potential with chiral helical symmetry.\cite{VS14,ME12,EAA13} The authors of Ref.\ \onlinecite{GJ13} studied the effect of strong SOI in a metallic substrate on the spin polarization of photoelectrons due to angular momentum selection.

A natural source of spin polarization is the SOI.\cite{Dresselhaus,Rashba,Winkler} On a one-dimensional (1D) wire, this interaction can be removed by a gauge transformation.\cite{GAM12,MH92,MJS02} The effect of SOI becomes non-trivial if one allows for quantum interference, via more than one electronic path. The simplest model of a spin filter is thus a two-path interferometer.\cite{MB04,CR06,HN07,AA11,MS13b} However, when the interferometer is connected to a single input and output 1D single-mode leads there is no spin polarization unless one breaks time-reversal symmetry.\cite{BCWJ97,KAA05} This fact stems from symmetry considerations of the scattering matrix,\cite{BCWJ97} which show that the combination of unitarity and time-reversal symmetry leads to a twofold degeneracy of transmission eigenvalues.\cite{BJH08,com} Due to this Kramers degeneracy of transmission eigenvalues, a finite spin polarization is forbidden in a two-terminal setup with single-mode leads.\cite{KAA05} Indeed, spin filtering does arise from the competition between the SOI and a magnetic Aharonov-Bohm flux.\cite{MB04,CR06,HN07,AA11,MS13b} Alternatively, it has been shown that spin polarization can be obtained in a two-terminal interferometer in the absence of magnetic flux, by assuming leakage of electrons from the interferometer to the environment, via side leads.\cite{MS13} As shown earlier, such a leakage breaks unitarity, and gives rise to complex site self-energies.\cite{AA02} Some of these physical sources for spin polarization have been used in the literature. Reference \onlinecite{GAM14} has introduced hopping between further neighbor sites along the helix, which gives rise to quantum interference. The recent observation of the CISS effect in protein-like helical molecules,\cite{MD13,EH15} as opposed to
ssDNA molecules, was attributed to the differences in such terms between these two systems.\cite{GAM14} Apparently, the hopping amplitudes in the protein-like helical molecules are larger than those in the ssDNA due to smaller distances between the nucleobases. References \onlinecite{GAM12,GAM14,VS14} also introduced phase-breaking processes (either via B\"{u}ttiker probes\cite{Buttiker} or via complex potentials in the Hamiltonian\cite{EKB95}). The latter processes may indeed result from the experimentally observed electron capture by DNA molecules.\cite{RSG05,MTZ13}

In this paper we present a simple tight-binding model for non-unitary electronic transport through a single helical molecule, which contains SOI on the nearest neighbor (NN) bonds
along the helix, and quantum interference, generated via hopping of electrons to the neighboring sites parallel to the helix axis. The unitarity is broken by allowing for leakage of electrons to side leads, which are connected to each site on the helix. We expect similar results for any model which generates complex site self-energies on the helix sites. The simplicity of the model allows an analytical solution, including a systematic analysis of the roles played by the various physical processes. In particular, we reproduce the experimentally observed increase of the relative outgoing spin polarization with the increase of the helix length, and find the direction of the maximal polarization.

The paper is organized as follows. Our model is defined in Sec.\ \ref{Sec 2A} and solved for the band structure of the infinitely long helix, assuming arbitrary SOI-induced spin rotation matrices within a unit cell in Sec.\ \ref{Sec 2B}. We then solve the scattering problem for a finite system (Sec.\ \ref{Sec 2C}) and introduce a complex site self-energy, due to leakage of electrons into absorbing channels (Sec.\ \ref{Sec Leakage}). The spin polarization of the transmitted electrons is studied in Sec.\ \ref{Sec 2D} for the general model defined in Secs.\ \ref{Sec 2A}-\ref{Sec 2C}. In Sec.\ \ref{Sec 2E} we introduce specific expressions for the SOI-induced spin rotation matrices corresponding to a helical geometry and study the resulting spin polarization in this case. We discuss and summarize the results in Sec.\ \ref{Sec 3}.
%
\section{The Model} \label{Sec 2A} 
\begin{figure}[ht!]
\begin{center}
\includegraphics[width=0.5\textwidth
]{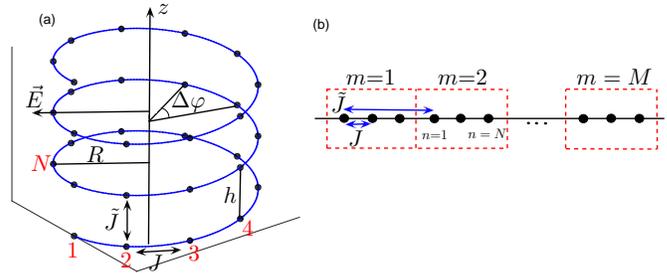}
\end{center}
\caption{\label{fig:Model}(Color online) Tight-binding model of a
single helical molecule with radius $R$, pitch $h$, and twist angle $\Delta\varphi$. Electrons can hop between adjacent sites
along the helix with hopping amplitude $J$ or vertically to the
$N$th neighbor with hopping amplitude $\tilde{J}$. Spin-orbit
interaction is assumed to act only between NN sites. (a) Schematic
view of the helical molecule. (b) Mapping of the model onto a
one-dimensional chain of $M$ unit cells, each containing $N$
sites.}
\end{figure}

To keep the model
analytically tractable, we assume that electrons can hop between
NN sites with hopping amplitude $J$ or along the axial direction,
to the $N$th neighbor, with a hopping amplitude $\tilde{J}$ [Fig.\
\ref{fig:Model}(a)]. The helical molecule can then be mapped onto
a one-dimensional chain of $M$ unit cells (labelled by $m=1, 2,
\ldots, M$), with a basis consisting of $N$ sites (labelled by
$n=1, 2, \ldots, N$), as shown in Fig.\ \ref{fig:Model}(b).

The Hamiltonian we study is
\begin{align}
\label{eq:1}&\mathcal{H}^{}_{mol}=\varepsilon^{}_{0}\sum^{M}_{m=1}\sum^{N}_{n=1}c^{\dag}_{m,n}c^{}_{m,n}\nonumber\\
&-\sum^{M}_{m=1}\sum^{N}_{n=1}\Big[Jc^{\dag}_{m,n+1}V^{}_{n}c^{}_{m,n}
+\tilde{J}c^{\dag}_{m+1,n}c^{}_{m,n}+\text{H.c.}\Big],
\end{align}
where
$c^{\dag}_{m,n}=(c^{\dag}_{m,n,\uparrow},c^{\dag}_{m,n,\downarrow})$
is the creation operator at site $m, n$ (with
$c^{\dag}_{m,N+1}=c^{\dag}_{m+1,1}$) and $V^{}_{n}$ is the unitary
matrix which describes the spin precession due to the SOI for an
electron hopping from site $m,n$ to site $m,n+1$. This matrix is
given by\cite{OY92}
\begin{align}
\label{eq:2}&V^{}_{n}=e^{i\mathbf{K}^{}_{n}\cdot\boldsymbol\sigma},
\end{align}
with $\boldsymbol\sigma$ being the vector of Pauli matrices and
\begin{align}
\label{eq:2b}&\mathbf{K}^{}_{n}=\lambda\,\mathbf{d}^{}_{n,n+1}\times\mathbf{E}^{}_{n,n+1},
\end{align}
where $\lambda$ is the parameter representing the SOI strength,
$\mathbf{d}^{}_{n,n+1}$ is the vector along the bond between site
$m,n$ and its NN site $m,n+1$ and $\mathbf{E}^{}_{n,n+1}$ is the
average electric field acting on an electron which moves along
this bond. For simplicity, we assume that the SOI affects only the
NN bonds and not the axial bonds. As shown below, a
spin-independent scalar hopping amplitude $\tilde{J}$ is
sufficient for the demonstration of the CISS
effect.\cite{Comment1}

In Secs.\ \ref{Sec 2B}-\ref{Sec 2D} we study the band structure of
an infinitely long chain and the scattering problem through a
finite chain assuming arbitrary unitary matrices $V^{}_{1},
\ldots, V^{}_{N}$. This allows for an arbitrary structure of the molecule within each unit cell.
This also makes the model general and applicable to
other systems different from helical chiral molecules, for
instance mesoscopic quantum networks. We then employ the specific
geometry of a helix in Sec.\ \ref{Sec 2E} to study the spin
polarization of the electrons transmitted through a single helical
chiral molecule.
\section{Band Structure} \label{Sec 2B}
To study the band structure of the chain we apply periodic
boundary conditions to the Hamiltonian \eqref{eq:1}, that is
$c^{\dag}_{m+M,n}=c^{\dag}_{m,n}$. The Hamiltonian \eqref{eq:1}
can then be diagonalized in the orbital space by first applying
the Fourier transform
\begin{align}
\label{eq:7}&c^{\dag}_{m,n}=\frac{1}{\sqrt{M}}\sum^{}_{k}c^{\dag}_{k,n}e^{-ikm},
\end{align}
where $k=2\pi r/M$ ($r=1, \ldots, M$) is the wave vector (in units
of $Na$, where $a$ is the distance between NN sites). Using also the identity $\sum^{M}_{m=1}e^{-i(k-k')m}=M\delta^{}_{k,k'}$,
the Hamiltonian \eqref{eq:1} then reads
\begin{align}
\label{eq:8}&\mathcal{H}^{}_{mol}=\sum^{N}_{n=1}\sum^{}_{k}\Bigl[\left(\varepsilon^{}_{0}-2\tilde{J}\cos k\right)c^{\dag}_{k,n}c^{}_{k,n}\nonumber\\
&-\Bigl(Jc^{\dag}_{k,n+1}V^{}_{n}c^{}_{k,n}+\text{H.c.}\Bigr)\Bigr].
\end{align}
Performing the gauge transformation
\begin{align}
\label{eq:9}&c^{\dag}_{k,n}=a^{\dag}_{k,n}e^{-ikn/N}\mathcal{V}^{n/N}V^{}_{N}V^{}_{N-1}\cdot\ldots\cdot
V^{}_{n},\nonumber\\
&\mathcal{V}\equiv V^{}_{N}V^{}_{N-1}\cdot\ldots\cdot V^{}_{1},
\end{align}
the Hamiltonian \eqref{eq:8} reduces to
\begin{align}
\label{eq:10}&\mathcal{H}^{}_{mol}=\sum^{}_{k}\sum^{N}_{n=1}\left[\varepsilon^{}_{0}-2\tilde{J}\cos k\right]a^{\dag}_{k,n}a^{}_{k,n}\nonumber\\
&-J\sum^{}_{k}\left[e^{-ik/N}\sum^{N}_{n=1}a^{\dag}_{k,n+1}\mathcal{V}^{1/N}a^{}_{k,n}+\text{H.c.}\right].
\end{align}
Applying a second Fourier transform,
\begin{align}
\label{eq:11}&a^{\dag}_{k,n}=\frac{1}{\sqrt{N}}\sum^{N}_{p=1}a^{\dag}_{k,p}e^{-2\pi
ipn/N},
\end{align}
and using the identity $\sum^{N}_{n=1}e^{-2\pi
i(p-p')n/N}=N\delta^{}_{p,p'}$, one finds
\begin{align}
\label{eq:12}&\mathcal{H}^{}_{mol}=\sum^{}_{k}\sum^{N}_{p=1}a^{\dag}_{k,p}H^{}_{p}\left(k\right)a^{}_{k,p},
\end{align}
where the spin space Hamiltonian for each band index $p$ is
\begin{align}
\label{eq:13}&H^{}_{p}\left(k\right)=\varepsilon^{}_{0}-2\tilde{J}\cos
k-J\Big[\mathcal{V}^{1/N}e^{-i\left(k+2\pi
p\right)/N}\nonumber\\
&+\text{H.c.}\Big].
\end{align}

\begin{figure}[ht!]
\begin{center}
\includegraphics[width=0.5\textwidth,height=0.32\textheight]{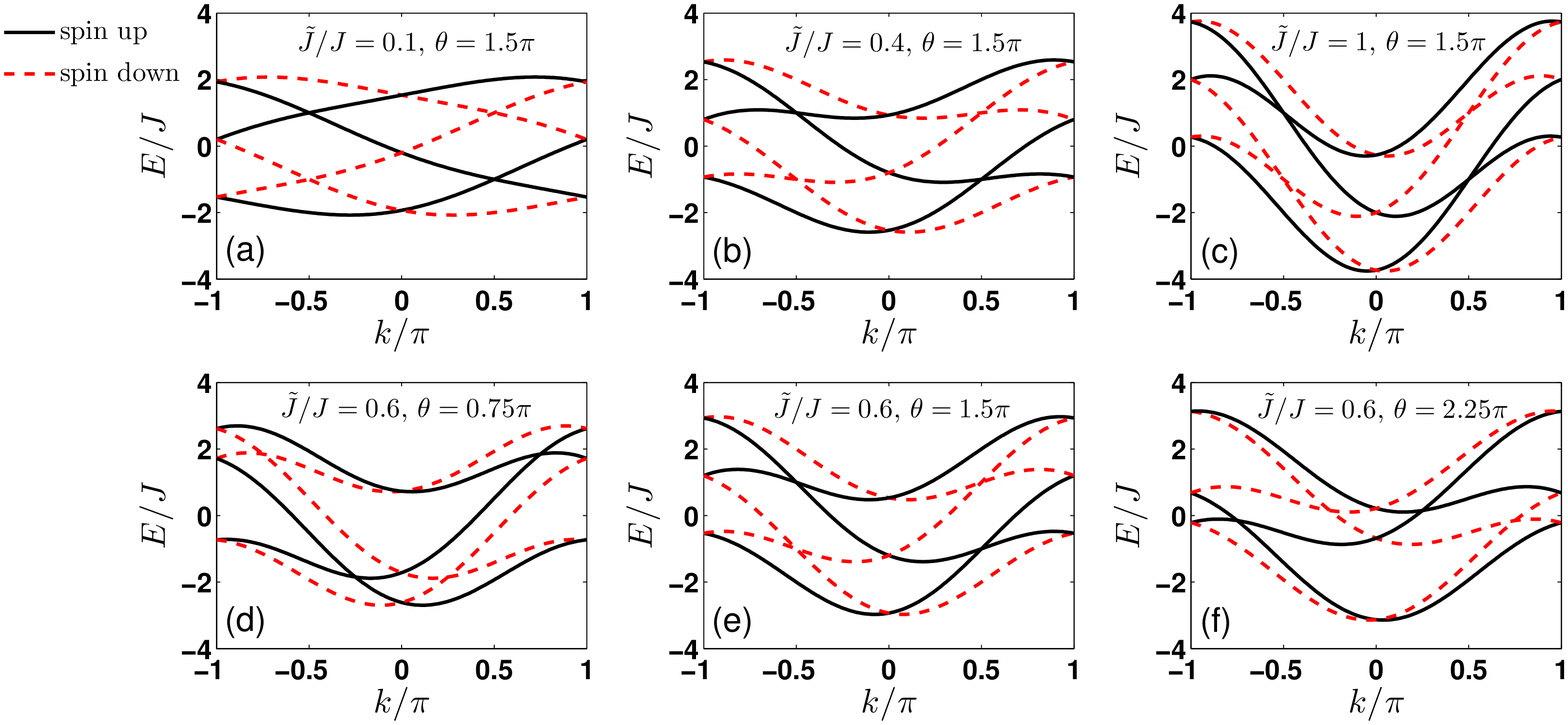}
\end{center}
\caption{\label{fig:band_structure}(Color online) Band structure of the infinite helix for $N=3$ sites in each unit cell. The site energy is $\varepsilon^{}_{0}=0$ and the values of $\tilde{J}/J$ and $\theta$ are specified in the legend of each panel. Solid (black) and dashed (red) bands correspond to spin up and down along the vector $\hat{\bold{n}}$, respectively.}
\end{figure}

Since $\mathcal{V}$ is a unitary matrix, it can always be written as
\begin{align}
\label{eq:14}&\mathcal{V}=e^{i\theta\,\hat{\bold{n}}\cdot\boldsymbol\sigma},
\end{align}
where $\hat{\bold{n}}$ is a unit vector and the angle $\theta$ is a measure of the strength $\lambda$ of the SOI. In the limit of vanishing SOI, $\lambda\rightarrow 0$, one has $V^{}_{n}\rightarrow\bold{1}$ for $n=1, \dots, N$ and therefore $\mathcal{V}\rightarrow\bold{1}$, or equivalently $\theta\rightarrow 0$. At small $\lambda$, an expansion of $\mathcal{V}$ shows that $\theta$ is linear in $\lambda$. In Sec.\ \ref{Sec 2E} we present a specific model of the helix, which gives an explicit dependence of $\theta$ on $\lambda$ (see Eqs.\ \eqref{eq:33} below). The matrix \eqref{eq:14} describes the spin precession of the electron after completing one turn along the helix (i.e. transport from site $m,n$ to site $m+1,n$). Upon completing one turn, the electron's spinor is rotated by an angle $2\theta$ about the direction $\hat{\bold{n}}$. The eigenvalues of the Hamiltonian \eqref{eq:13} then read
\begin{align}
\label{eq:15}&E^{}_{p,\sigma}(k)=\varepsilon^{}_{0}-2\tilde{J}\cos
k-2J\cos\left(\frac{k+2\pi p-\sigma\theta}{N}\right),
\end{align}
where $\sigma=\pm 1$ corresponds to the eigenspinors $\ket{\pm\hat{\bold{n}}}$ of the spin projection along $\hat{\bold{n}}$, i.e.\ $\hat{\bold{n}}\cdot\boldsymbol\sigma\ket{\sigma\hat{\bold{n}}}=\sigma\ket{\sigma\hat{\bold{n}}}$. With $M\gg 1$, $k$ becomes quasi-continuous and the band structure consists of $2N$ spin-resolved bands. The spin degeneracy has been lifted by the SOI but time-reversal symmetry is still conserved. As a result, the bands described by Eq.\ \eqref{eq:15} are Kramers
degenerate, satisfying $E^{}_{p,\sigma}(k)=E^{}_{-p,-\sigma}(-k)$. The band structure \eqref{eq:15} is plotted in Fig.\ \ref{fig:band_structure} with $k$ lying within the first Brillouin zone ($-\pi<k\leq\pi$), for $\varepsilon^{}_{0}=0$ and $N=3$ and for various values of $\tilde{J}/J$ and $\theta$.

For a given energy $E$ in the left-hand side of Eq.\ \eqref{eq:15}, the solutions with real (complex) wave vector $k$ describe propagating (evanescent) waves. In an infinite chain one
has to consider only the propagating solutions. The number of propagating waves for a given energy can be deduced graphically by considering the number of crossings of a horizontal line (corresponding to the given energy) in the band structure (Fig.\ \ref{fig:band_structure}). As seen in Fig.\ \ref{fig:band_structure}, at a given energy $E$ upward propagating
solutions ($v=dE/dk>0$) come in pairs of opposite spins $\ket{\pm\hat{\bold{n}}}$. These spins are independent of the wave vector $k$ and the band index $p$. As a result, conduction
electrons occupy states below the Fermi energy in pairs of opposite spins. The total spin cancels out completely, leading to zero spin polarization of the transmitted electrons.

The arguments presented above are valid for an infinitely long
chain. To study transport through a helical molecule  in more
detail (and to compare with the experiments) requires the solution of a scattering off a finite helix.
\section{Scattering problem} \label{Sec 2C}
We assume that the chain is connected to two ideal semi-infinite
one-dimensional leads with hopping amplitude $J^{}_{0}$, free of
SOI, at the left ($l\leq 1$) and right ($l\geq NM$) edges of Fig.\ \ref{fig:Model}(b).
Consider now a scattering state at a given energy $E$,
\begin{align}
\label{eq:leads}&\ket{\psi^{}_{l}}=\begin{cases}\ket{\chi^{}_{in}}e^{ik^{}_{0}\left(l-1\right)}+r\ket{\chi^{}_{r}}e^{-ik^{}_{0}\left(l-1\right)}
& \quad l\leq 1 \\
t\ket{\chi^{}_{t}}e^{ik^{}_{0}\left(l-NM\right)} & \quad l\geq NM,
\end{cases}
\end{align}
with $\ket{\chi^{}_{in}}$, $\ket{\chi^{}_{r}}$ and
$\ket{\chi^{}_{t}}$ being the normalized incoming, reflected and
transmitted spinors, respectively, and the wave vector $k^{}_{0}$
(in units of the lattice constant of the leads) satisfying the
dispersion relation $E=-2J^{}_{0}\cos k^{}_{0}$.

The solution of the scattering problem is obtained by solving the
tight-binding Schr\"{o}dinger equations for the spinors
$\ket{\psi^{}_{m,n}}$ ($1\leq n\leq N$, $1\leq m\leq M$). For $l=1$ and
$l=NM$ the solutions inside and outside the chain coincide
provided that
\begin{align}
\label{eq:23}&\ket{\chi^{}_{in}}+r\ket{\chi^{}_{r}}=\ket{\psi^{}_{1,1}},\nonumber\\
&t\ket{\chi^{}_{t}}=\ket{\psi^{}_{M,N}}.
\end{align}

The equations inside the chain ($1\leq n\leq N$, $2\leq m\leq
M-1$) read
\begin{widetext}
\begin{align}
\label{eq:17}&\left(E-\varepsilon^{}_{0}\right)\ket{\psi^{}_{m,1}}=-JV^{}_{N}\ket{\psi^{}_{m-1,N}}-JV^{\dag}_{1}\ket{\psi^{}_{m,2}}-\tilde{J}\left(\ket{\psi^{}_{m-1,1}}+\ket{\psi^{}_{m+1,1}}\right),\nonumber\\
&\left(E-\varepsilon^{}_{0}\right)\ket{\psi^{}_{m,n}}=-JV^{}_{n-1}\ket{\psi^{}_{m,n-1}}-JV^{\dag}_{n}\ket{\psi^{}_{m,n+1}}-\tilde{J}\left(\ket{\psi^{}_{m-1,n}}+\ket{\psi^{}_{m+1,n}}\right), \qquad 2\leq n\leq N-1,\nonumber\\
&\left(E-\varepsilon^{}_{0}\right)\ket{\psi^{}_{m,N}}=-JV^{}_{N-1}\ket{\psi^{}_{m,N-1}}-JV^{\dag}_{N}\ket{\psi^{}_{m+1,1}}-\tilde{J}\left(\ket{\psi^{}_{m-1,N}}+\ket{\psi^{}_{m+1,N}}\right).
\end{align}
Similarly, the corresponding tight-binding equations for the $N$
sites in the unit cell $m=1$ are\cite{Comment2}
\begin{align}
\label{eq:24}&\left(E-y^{}_{0}\right)\ket{\psi^{}_{1,1}}=2iJ^{}_{0}\sin k^{}_{0}\ket{\chi^{}_{in}}-JV^{\dag}_{1}\ket{\psi^{}_{1,2}}-\tilde{J}\ket{\psi^{}_{2,1}},\nonumber\\
&\left(E-\varepsilon^{}_{0}\right)\ket{\psi^{}_{1,n}}=-JV^{}_{n-1}\ket{\psi^{}_{1,n-1}}-JV^{\dag}_{n}\ket{\psi^{}_{1,n+1}}-\tilde{J}\ket{\psi^{}_{2,n}}, \qquad 2\leq n\leq N-1,\nonumber\\
&\left(E-\varepsilon^{}_{0}\right)\ket{\psi^{}_{1,N}}=-JV^{}_{N-1}\ket{\psi^{}_{1,N-1}}-JV^{\dag}_{N}\ket{\psi^{}_{2,1}}-\tilde{J}\ket{\psi^{}_{2,N}},
\end{align}
whereas those in the unit cell $m=M$ are
\begin{align}
\label{eq:24b}&\left(E-\varepsilon^{}_{0}\right)\ket{\psi^{}_{M,1}}=-JV^{\dag}_{1}\ket{\psi^{}_{M,2}}-JV^{}_{N}\ket{\psi^{}_{M-1,N}}-\tilde{J}\ket{\psi^{}_{M-1,1}},\nonumber\\
&\left(E-\varepsilon^{}_{0}\right)\ket{\psi^{}_{M,n}}=-JV^{}_{n-1}\ket{\psi^{}_{M,n-1}}-JV^{\dag}_{n}\ket{\psi^{}_{M,n+1}}-\tilde{J}\ket{\psi^{}_{M-1,n}}, \qquad 2\leq n\leq N-1,\nonumber\\
&\left(E-y^{}_{0}\right)\ket{\psi^{}_{M,N}}=-JV^{}_{N-1}\ket{\psi^{}_{M,N-1}}-\tilde{J}\ket{\psi^{}_{M-1,N}}.
\end{align}
Here $y^{}_{0}=\varepsilon^{}_{0}-J^{}_{0}e^{ik^{}_{0}}$ and we
substituted
$r\ket{\chi^{}_{r}}=-\ket{\chi^{}_{in}}+\ket{\psi^{}_{1,1}}$ and
$t\ket{\chi^{}_{t}}=\ket{\psi^{}_{M,N}}$ from Eqs.\ \eqref{eq:23}.

To solve Eqs.\ \eqref{eq:17} we apply the same transformation as
in Sec.\ \ref{Sec 2C} [see Eqs.\ \eqref{eq:7} and \eqref{eq:9}],
that is
\begin{align}
\label{eq:18}&\ket{\psi^{}_{m,n}}=e^{ikm}e^{ikn/N}V^{\dag}_{n}\cdot\ldots\cdot
V^{\dag}_{N}\mathcal{V}^{-n/N}\ket{\varphi^{}_{n}}.
\end{align}
The wave vector $k$ will be determined below. Equations
\eqref{eq:17}-\eqref{eq:24b} then reduce to
\begin{align}
\label{eq:19}&\left(E-\varepsilon^{}_{0}+2\tilde{J}\cos k\right)\ket{\varphi^{}_{1}}=-Je^{-ik/N}\mathcal{V}^{1/N}\ket{\varphi^{}_{N}}-Je^{ik/N}\mathcal{V}^{-1/N}\ket{\varphi^{}_{2}},\nonumber\\
&\left(E-\varepsilon^{}_{0}+2\tilde{J}\cos
k\right)\ket{\varphi^{}_{n}}=-Je^{-ik/N}\mathcal{V}^{1/N}\ket{\varphi^{}_{n-1}}-Je^{ik/N}\mathcal{V}^{-1/N}\ket{\varphi^{}_{n+1}}, \qquad 2\leq n\leq N-1,\nonumber\\
&\left(E-\varepsilon^{}_{0}+2\tilde{J}\cos
k\right)\ket{\varphi^{}_{N}}=-Je^{-ik/N}\mathcal{V}^{1/N}\ket{\varphi^{}_{N-1}}-Je^{ik/N}\mathcal{V}^{-1/N}\ket{\varphi^{}_{1}}.
\end{align}
\begin{align}
\label{eq:19b}&\left(E-y^{}_{0}+\tilde{J}e^{ik}\right)e^{ik\left(1+1/N\right)}\mathcal{V}^{-1/N}\ket{\varphi^{}_{1}}+Je^{ik\left(1+2/N\right)}\mathcal{V}^{-2/N}\ket{\varphi^{}_{2}}=2iJ^{}_{0}\sin k^{}_{0}\mathcal{V}\ket{\chi^{}_{in}},\nonumber\\
&
\left(E-\varepsilon^{}_{0}+\tilde{J}e^{ik}\right)\ket{\varphi^{}_{n}}+Je^{-ik/N}\mathcal{V}^{1/N}\ket{\varphi^{}_{n-1}}
+Je^{ik/N}\mathcal{V}^{-1/N}\ket{\varphi^{}_{n+1}}
=0, \qquad 2\leq n\leq N-1,\nonumber\\
&
\left(E-\varepsilon^{}_{0}+\tilde{J}e^{ik}\right)\ket{\varphi^{}_{N}}
+Je^{-ik/N}\mathcal{V}^{1/N}\ket{\varphi^{}_{N-1}}+Je^{ik/N}\mathcal{V}^{-1/N}\ket{\varphi^{}_{1}}
=0.
\end{align}
\begin{align}
\label{eq:19c}&
\left(E-\varepsilon^{}_{0}+\tilde{J}e^{-ik}\right)\ket{\varphi^{}_{1}}+Je^{-ik/N}\mathcal{V}^{1/N}\ket{\varphi^{}_{N}}
+Je^{ik/N}\mathcal{V}^{-1/N}\ket{\varphi^{}_{2}}
=0,\nonumber\\
&
\left(E-\varepsilon^{}_{0}+\tilde{J}e^{-ik}\right)\ket{\varphi^{}_{n}}
+Je^{-ik/N}\mathcal{V}^{1/N}\ket{\varphi^{}_{n-1}}+Je^{ik/N}\mathcal{V}^{-1/N}\ket{\varphi^{}_{n+1}}
=0, \qquad 2\leq n\leq N-1,\nonumber\\
&
\left(E-y^{}_{0}+\tilde{J}e^{-ik}\right)\ket{\varphi^{}_{N}}+Je^{-ik/N}\mathcal{V}^{1/N}\ket{\varphi^{}_{N-1}}
=0.
\end{align}
\end{widetext}
Looking at Eqs.\ \eqref{eq:19}-\eqref{eq:19c}, one sees that they
become simple if we set the $\ket{\varphi^{}_{n}}$'s to be eigenstates
of $\mathcal{V}$. In particular, the solution of the eigenvalue
problem \eqref{eq:19} gives the same band structure
$E^{}_{p,\sigma}(k)$ ($\sigma=\pm 1$ is a spin index and $p=1, 2,
\dots, N$ is a band index) as in Eq.\ \eqref{eq:15} and the
corresponding eigenspinors
\begin{align}
\label{eq:20}&\ket{\varphi^{}_{n,p,\sigma}}=e^{2\pi
inp/N}\ket{\sigma\hat{\bold{n}}}.
\end{align}
At a given energy $E$, the dispersion relations \eqref{eq:15} can
be written as
\begin{align}
\label{eq:poly}&E-\varepsilon^{}_{0}+2\tilde{J}\cos\left(Ny\right)+2J\cos\left(\frac{\theta}{N}\right)\cos
y=\nonumber\\
&-2J\sigma\sin\left(\frac{\theta}{N}\right)\sin y,
\end{align}
where $y=\left(k+2\pi p\right)/N$. The left-hand side is a
polynomial of degree $N$ in $x=\cos y$. Squaring the two sides of
this equation gives a polynomial of degree $2N$ in $x$, yielding
$2N$ solutions $\{x^{}_{j}=\cos y^{}_{j}, \: j=1, 2, \ldots,
2N\}$. For a specific value of $\sigma$, substituting each of
these solutions into Eq.\ \eqref{eq:poly} identifies $\sin
y^{}_{j,\sigma}$, and hence $e^{iy^{}_{j,\sigma}}=x^{}_{j}+i\sin
y^{}_{j,\sigma}$. For each energy $E$ and spin $\sigma$, the solutions of Eq.\
\eqref{eq:17}-\eqref{eq:24b} must therefore be  linear
combinations of the corresponding $2N$ solutions,
\begin{align}
\label{eq:psimn}&\ket{\psi^{}_{m,n,\sigma}}=e^{-i\sigma\theta
n/N}V^{\dag}_{n}\ldots V^{\dag}_{N}\sum_{j=1}^{2N}A^{\sigma}_{j}
e^{iy^{}_{j,\sigma}\left(mN+n\right)}\ket{\sigma\hat{\bf n}}.
\end{align}
The amplitudes $A^{\sigma}_{j}$ are determined by Eqs.\
\eqref{eq:24} and \eqref{eq:24b}. If
$\ket{\chi^{}_{in}}=\ket{\hat{\bold{n}}}$, then one finds that
$A^{-}_{j}=0$, and all the spinors $\ket{\psi^{}_{m,n}}$ have only
$\sigma=1$. Similarly, if
$\ket{\chi^{}_{in}}=\ket{-\hat{\bold{n}}}$, one has $A^{+}_{j}=0$,
and all the spinors $\ket{\psi^{}_{m,n}}$ have only $\sigma=-1$.
The remaining amplitudes are found by solving the $2N$ linear
equations
\begin{widetext}
\begin{align}
\label{eq:amplitudes1}&\sum^{2N}_{j=1}\left[E-y^{}_{0}+Je^{i\left(y^{}_{j,\sigma}-\sigma\theta/N\right)}+\tilde{J}e^{iy^{}_{j,\sigma}N}\right]e^{i\left[y^{}_{j,\sigma}\left(N+1\right)-\sigma\theta/N\right]}A^{\sigma}_{j}=2iJ^{}_{0}\sin k^{}_{0}e^{i\sigma\theta},\nonumber\\
&\sum^{2N}_{j=1}\left[E-\varepsilon^{}_{0}+2J\cos\left(y^{}_{j,\sigma}-\sigma\theta/N\right)+\tilde{J}e^{iy^{}_{j,\sigma}N}\right]e^{iy^{}_{j,\sigma}\left(N+n\right)}A^{\sigma}_{j}=0, \qquad 2\leq n\leq N,\nonumber\\
&\sum^{2N}_{j=1}\left[E-\varepsilon^{}_{0}+2J\cos\left(y^{}_{j,\sigma}-\sigma\theta/N\right)+\tilde{J}e^{-iy^{}_{j,\sigma}N}\right]e^{iy^{}_{j,\sigma}\left(MN+n\right)}A^{\sigma}_{j}=0, \qquad 1\leq n\leq N-1,\nonumber\\
&\sum^{2N}_{j=1}\left[E-y^{}_{0}+Je^{-i\left(y^{}_{j,\sigma}-\sigma\theta/N\right)}+\tilde{J}e^{-iy^{}_{j,\sigma}N}\right]e^{iy^{}_{j,\sigma}N\left(M+1\right)}A^{\sigma}_{j}=0.
\end{align}
\end{widetext}
Using Eq.\ \eqref{eq:23}, it then becomes clear that for
$\ket{\chi^{}_{in}}=\ket{\hat{\bold{n}}}$
($\ket{\chi^{}_{in}}=\ket{-\hat{\bold{n}}}$) the reflected
electron is also polarized along $\hat{\bf n}$ ($-\hat{\bf n}$),
while the transmitted electron is described by
$t\ket{\chi^{}_{t}}=\ket{\psi^{}_{M,N,\sigma}}=V^{\dag}_{N}\sum^{2N}_{j=1}A^{\sigma}_{j}e^{i\left[y^{}_{j,\sigma}N\left(M+1\right)-\sigma\theta\right]}\ket{\sigma\hat{\bold{n}}}$,
and thus $\ket{\chi^{}_{t}}=\ket{\sigma\hat{\bold{n}}'}$, where
\begin{align}
\label{eq:27}&\ket{\sigma\hat{\bold{n}}'}\equiv
V^{\dag}_{N}\ket{\sigma\hat{\bold{n}}}.
\end{align}

In the general case, it is convenient to write
$\ket{\chi^{}_{in}}$ in the basis $\ket{\pm\hat{\bold{n}}}$ and
conclude that $r\ket{\chi^{}_{r}}=\mathcal{R}\ket{\chi^{}_{in}}$
and $t\ket{\chi^{}_{t}}=\mathcal{T}\ket{\chi^{}_{in}}$, with the
$2\times 2$ reflection and transmission amplitude matrices
\begin{align}
\label{eq:25}&\mathcal{R}=r^{}_{\uparrow\uparrow}\ket{\hat{\bold{n}}}\bra{\hat{\bold{n}}}+r^{}_{\downarrow\downarrow}\ket{-\hat{\bold{n}}}\bra{-\hat{\bold{n}}},\nonumber\\
&\mathcal{T}=t^{}_{\uparrow\uparrow}\ket{\hat{\bold{n}}'}\bra{\hat{\bold{n}}}+t^{}_{\downarrow\downarrow}\ket{-\hat{\bold{n}}'}\bra{-\hat{\bold{n}}},
\end{align}
with the the reflection and transmission amplitudes
\begin{align}
\label{eq:26}&r^{}_{\sigma\sigma}=-1+e^{-i\sigma\theta\left(1+1/N\right)}\sum^{2N}_{j=1}A^{\sigma}_{j}e^{iy^{}_{j,\sigma}\left(N+1\right)},\nonumber\\
&t^{}_{\sigma\sigma}=e^{-i\sigma\theta}\sum^{2N}_{j=1}A^{\sigma}_{j}e^{iy^{}_{j,\sigma}N\left(M+1\right)}.
\end{align}

From now on, let us concentrate on the case $N=2$ and demonstrate
the above procedure for this simple case. For a given energy $E$,
the dispersion relations \eqref{eq:15} yield the wave vectors
inside the chain as the solutions of a quartic equation in
$x\equiv\cos\left(k/2+\pi p\right)$,
\begin{align}
\label{eq:21}&4\tilde{J}^{2}x^{4}+4J\tilde{J}\cos\left(\frac{\theta}{2}\right)x^{3}+\left[2\tilde{J}\left(E-\varepsilon^{}_{0}-2\tilde{J}\right)+J^{2}\right]x^{2}\nonumber\\
&+\left(E-\varepsilon^{}_{0}-2\tilde{J}\right)J\cos\left(\frac{\theta}{2}\right)x+\frac{1}{4}\left(E-\varepsilon^{}_{0}-2\tilde{J}\right)^{2}\nonumber\\
&-J^{2}\sin^{2}\left(\frac{\theta}{2}\right)=0.
\end{align}
This equation has four solutions for $x$. Each solution $x^{}_{j}$
($j=1, 2, 3, 4$) leads to two opposite values of $y^{}_{j,\sigma}$,
corresponding to waves propagating in opposite directions and
carrying opposite spins (this follows from time-reversal symmetry,
as discussed above).
\section{Leakage} \label{Sec Leakage}
For real values of $\varepsilon^{}_{0}$, some solutions of Eq.\ \eqref{eq:21} correspond to real values of the wave vector $k=2y$. These propagating wave solutions yield four energy bands (two bands for each spin), similar to those shown in Fig.\ \ref{fig:band_structure}. At each energy $E$, these waves come in pairs of opposite spins and therefore do not give rise to spin splitting. For energies in the gaps between these bands, the solutions for $y$ become complex, and the corresponding waves become evanescent. The imaginary part of $y$ corresponds to the inverse decay length of the wave function. For real values of $\varepsilon^{}_{0}$, the coefficients of the polynomial equation \eqref{eq:21} are real. Therefore, if $y$ is a complex solution, then its complex conjugate $y^{\ast}$ is also a solution. Combining this with the time-reversal symmetry, it follows that evanescent solutions also come in pairs of opposite spins with the same decay lengths. Consequently, the net spin polarization of the scattered electrons still vanishes. This agrees with symmetry considerations of the scattering matrix,\cite{BCWJ97,BJH08} which show that the combination of unitarity and time-reversal symmetry forbids a finite spin polarization in a two-terminal setup with single-mode leads.\cite{KAA05}

In this paper we consider the possibility of generating a finite spin polarization in a two-terminal setup by breaking the unitarity of the scattering matrix. Specifically, we adopt the same approach as in Refs.\ \onlinecite{MS13} and \onlinecite{AA02}, and consider the leakage of electrons out of the system. Each site on the helix is connected to an absorbing
channel, modeled as a one-dimensional tight-binding chain (free of SOI) whose site energies are set to zero. The hopping amplitude on the first bond of each absorbing channel is $J^{}_{x}$, whereas the other bonds have a hopping amplitude $J^{}_{0}$. By assuming only outgoing waves on these absorbing channels, it was shown in Refs.\ \onlinecite{MS13} and \onlinecite{AA02} that the sole effect of these channels is to introduce a complex site self-energy for each site of the original chain, i.e.\ $\varepsilon^{}_{0}\rightarrow\tilde{\varepsilon}^{}_{0}$ with
\begin{align}
\label{eq:16}&\tilde{\varepsilon}^{}_{0}=\varepsilon^{}_{0}-\frac{|J^{}_{x}|^2e^{iq}}{J^{}_{0}},
\end{align}
where $q$ is the wave vector (in units of the lattice constant of the leads) of the outgoing waves in the absorbing channels, determined by the dispersion relation $E=-2J^{}_{0}\cos q$.

With the leakage, $\tilde{\varepsilon}^{}_{0}$ is always complex, and all the solutions for the wave vectors $Ny$ acquire finite imaginary parts. In contrast to the unitary case (real
$\tilde{\varepsilon}^{}_{0}$),  the evanescent waves associated with opposite spins now have different decay lengths. This opens up the possibility for a finite spin polarization, as discussed in detail below. It is important to note that one has to consider interference terms (represented by $\tilde{J}$ in our model) in order to achieve a finite spin polarization, even if $\tilde{\varepsilon}^{}_{0}$ is complex. As discussed in the introduction, the SOI can be gauged out if $\tilde{J}=0$, resulting in a trivial problem with zero spin polarization. The technical reason for this can also be seen from Eq.\ \eqref{eq:15}. With $\tilde{J}=0$, the solutions for $y$ become $y^{}_{\pm,\sigma}=\sigma\theta/N\pm\arccos\left[\left(\tilde{\varepsilon}^{}_{0}-E\right)/2J\right]$, so that the solutions with $\sigma=1$ have the same decay lengths as those with $\sigma=-1$, and no net spin polarization appears.

In the next sections we study quantitatively the spin polarization for $N=2$ and identify the direction of maximum polarization $\hat{\bold{n}}'$ for an arbitrary value of $N$ using Eq.\ \eqref{eq:27}. We show that some important conclusions regarding the direction of the spin polarization of the transmitted electrons can be obtained for an arbitrary value of $N$. This includes, for example, the reversal of the spin polarization along the $z$-axis with reversal of the helix chirality, as found experimentally (see Sec.\ \ref{Sec 2E}).
\section{Spin Polarization} \label{Sec 2D}
\begin{figure}[ht!]
\begin{center}
\subfigure[\label{fig:finite_polarization_E}]{\label{fig:finite_polarization_E}
\includegraphics[width=0.5\textwidth,height=0.24\textheight]{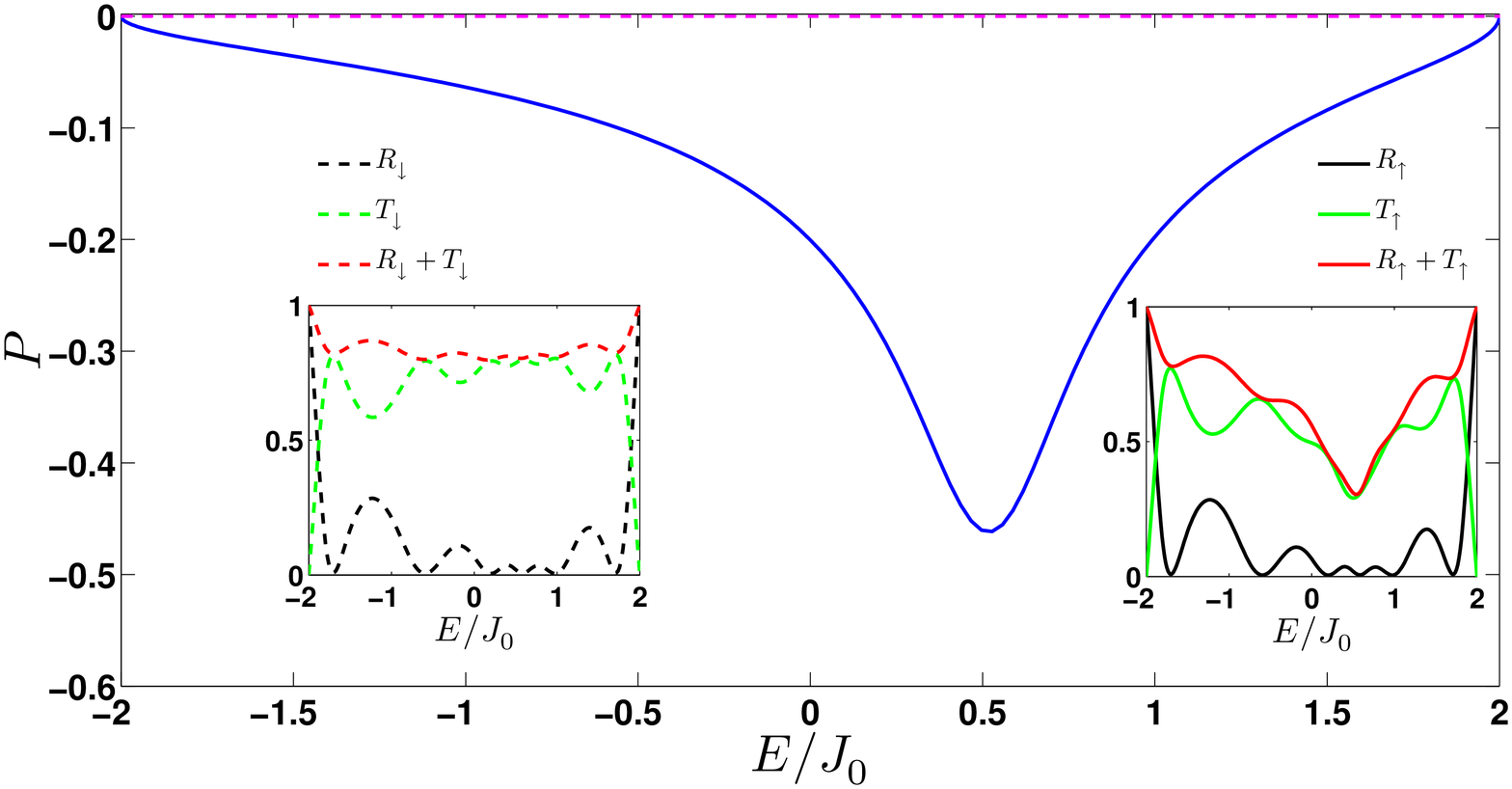}}
\subfigure[\label{fig:finite_polarization_SOI}]{\label{fig:finite_polarization_SOI}
\includegraphics[width=0.5\textwidth,height=0.24\textheight]{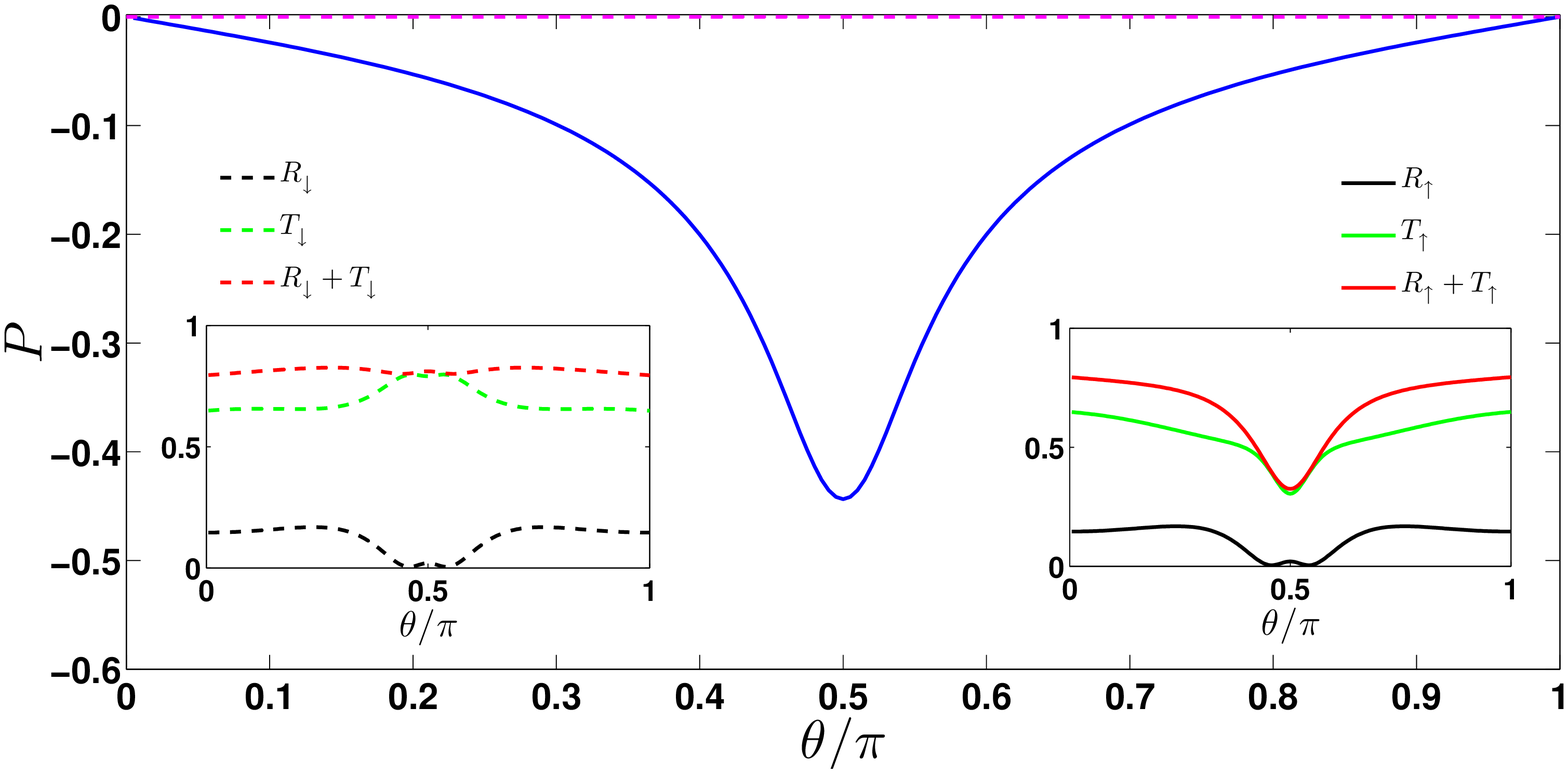}}
\end{center}
\caption{\label{fig:finite_polarization}(Color online) Spin polarization (solid blue) in a chain of $M=6$ unit cells (other parameters are $N=2$, $J=1.5J^{}_{0}$, $\tilde{J}=0.6J^{}_{0}$ and
$J^{}_{x}=0.2J^{}_{0}$) (a) as a function of energy (in units of $J^{}_{0}$) with $\theta=0.4\pi$ and (b) as a function of $\theta$ with $E=0$ (center of the band). The spin polarization vanishes (dashed magenta) if either $J^{}_{x}=0$ (unitary chain), $\tilde{J}=0$ (NN chain) or $\theta=0$ (no SOI). Inset: reflection (black) and transmission (green) coefficients for spin up (solid) and spin down (dashed). The red line is the sum these coefficients.}
\end{figure}

\begin{figure}[ht!]
\begin{center}
\includegraphics[width=0.5\textwidth,height=0.24\textheight]{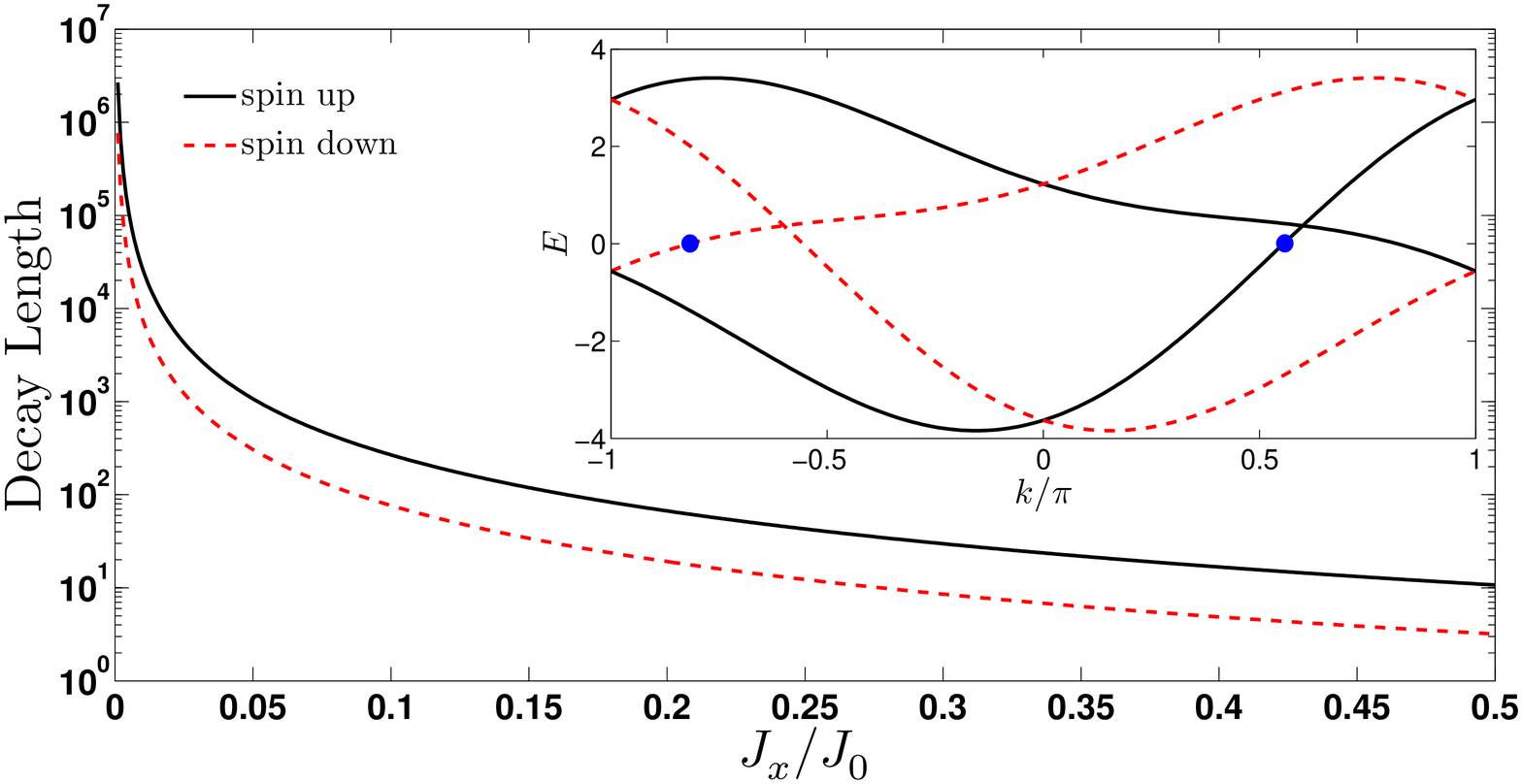}
\end{center}
\caption{\label{fig:decay_length}(Color online) Decay length of a pair of spin up (solid black) and spin down (dashed red) as function of $J^{}_{x}/J^{}_{0}$ for $N=2$, $J=1.5J^{}_{0}$, $\tilde{J}=0.6J^{}_{0}$, $\theta=0.4\pi$ and $E=0$. The inset shows the band structure corresponding to the chosen parameters and the pair of opposite spins propagating upward inside the chain ($v=dE/dk>0$) at energy $E=0$ is indicated by blue points.}
\end{figure}

Equation \eqref{eq:25} shows that for an incoming electron polarized along $\pm\hat{\bold{n}}$, the reflected and transmitted waves are polarized along $\pm\hat{\bold{n}}$ and
$\pm\hat{\bold{n}}'$, respectively. For an unpolarized incident beam, we use the second of Eqs.\ \eqref{eq:25} to write the polarization of the outgoing beam along $\hat{\bold{n}}'$ as
\begin{align}
\label{eq:28}&P^{}_{\hat{\bold{n}}'}\equiv\frac{\Tr\left[\mathcal{T}^{\dag}\left(\hat{\bold{n}}'\cdot\boldsymbol\sigma\right)\mathcal{T}\right]}{\Tr\left[\mathcal{T}^{\dag}\mathcal{T}\right]}=\frac{|t^{}_{\uparrow\uparrow}|^{2}-|t^{}_{\downarrow\downarrow}|^{2}}{|t^{}_{\uparrow\uparrow}|^{2}+|t^{}_{\downarrow\downarrow}|^{2}}.
\end{align}
Figure \ref{fig:finite_polarization} shows the spin polarization along the direction $\hat{\bold{n}}'$ [Eq.\ \eqref{eq:28}] and the reflection and transmission coefficients for an incoming spin up ($R^{}_{\uparrow}\equiv|r^{}_{\uparrow\uparrow}|^{2}$ and $T^{}_{\uparrow}\equiv|t^{}_{\uparrow\uparrow}|^{2}$) and spin down ($R^{}_{\downarrow}\equiv|r^{}_{\downarrow\downarrow}|^{2}$ and $T^{}_{\downarrow}\equiv|t^{}_{\downarrow\downarrow}|^{2}$) along the direction $\hat{\bold{n}}$. No spin polarization is achieved when either $J^{}_{x}=0$ or $\tilde{J}=0$. A finite value of $\tilde{J}$ gives rise to quantum interference whereas a finite $J^{}_{x}$ breaks unitarity and yields a complex site self-energy, and therefore different decay lengths for upward propagating waves with opposite spins. Figure\ \ref{fig:decay_length} shows the decay lengths (the inverse of the imaginary part of the wave vector) in units of the lattice constant as function of $J^{}_{x}$ (in units of $J^{}_{0}$) for a situation in which a pair of opposite spins are propagating upward ($v=dE/dk>0$) inside the chain (the parameters are the same as in Fig.\ \ref{fig:finite_polarization}). The decay lengths of both waves decrease when $J^{}_{x}$ increases but have a different value for each spin.

\begin{figure}[ht!]
\begin{center}
\includegraphics[width=0.5\textwidth,height=0.24\textheight]{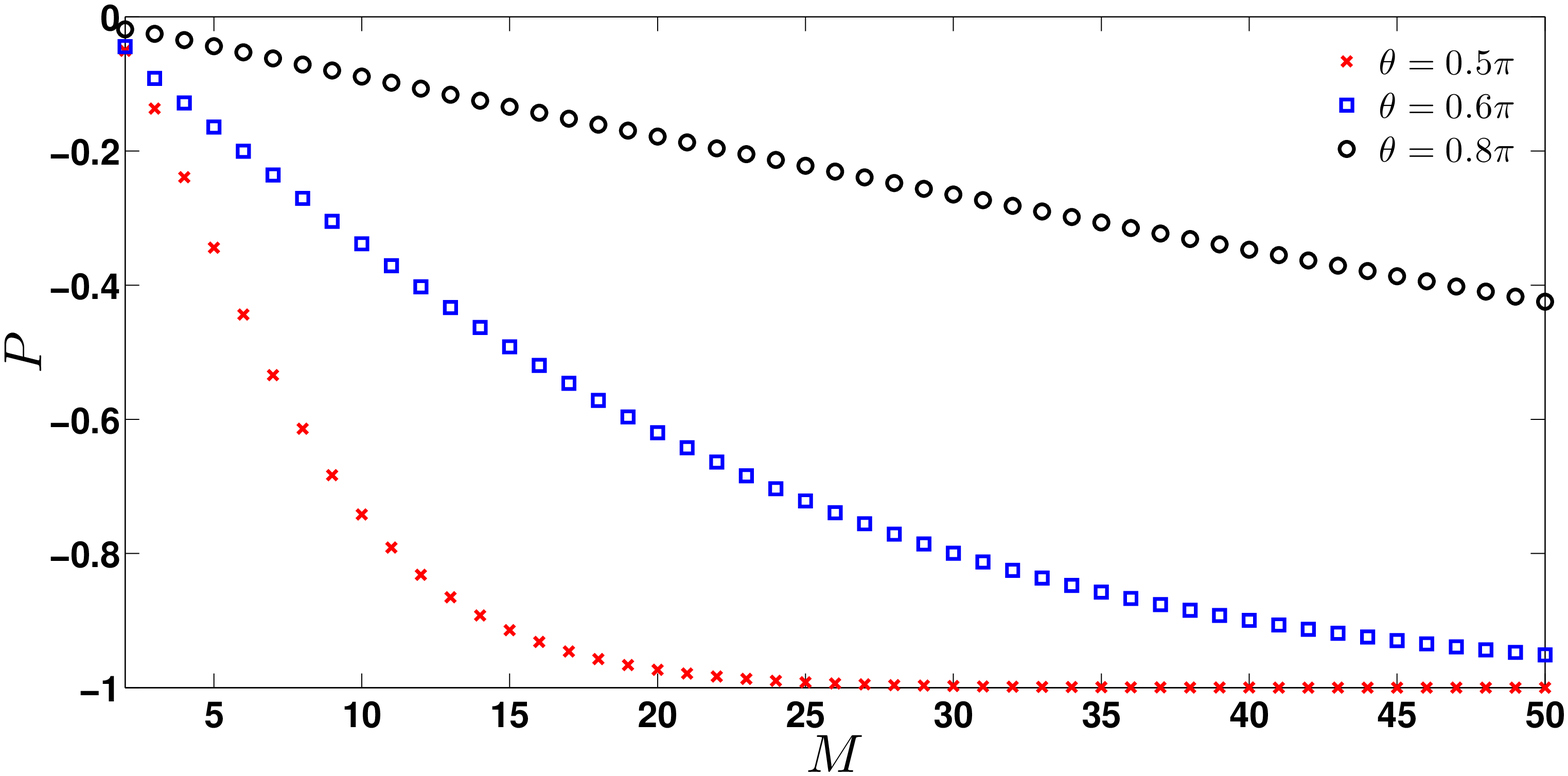}
\end{center}
\caption{\label{fig:length_dependence}(Color online) Spin polarization along $\hat{\bold{n}}'$ as a function of the number of unit cells $M$ for various values of $\theta$, with $E=0$, $J=1.5J^{}_{0}$, $\tilde{J}=0.6J^{}_{0}$ and $J^{}_{x}=0.2J^{}_{0}$.}
\end{figure}

In our non-unitary model the total transmission decreases with increasing length of the molecule. However, since the decay is larger for one of the spin components, the polarization is expected to increase with increasing length of the molecule, as shown in Fig.\ \ref{fig:length_dependence}. Both these conclusions are in agreement with the experimental observations.\cite{GB11,XZ11}

The results in Figs.\ \ref{fig:finite_polarization} and \ref{fig:length_dependence} depend on the hopping amplitudes $J,\ \tilde{J}$ and $J^{}_x$. When we set all of these amplitudes to be equal to $J^{}_0$ we find much larger values of $P$, at smaller values of $\theta$. For $M=6$, we find $|P|\approx 0.4$ at $\theta=0.05\pi$ and $|P|\approx 0.8$ at $\theta=0.15\pi$. Since none of these parameters are known for specific experimental systems, it is not easy to estimate values of $\theta$ which are needed to achieve a significant polarization. We return to this issue in the next section.

Our simple model allows studying the polarization along an arbitrary direction $\hat{\bold{m}}=\left(\cos\delta\sin\gamma,\sin\delta\sin\gamma,\cos\gamma\right)$. With
$\ket{\hat{\bold{n}}'}=\cos\left(\frac{\alpha}{2}\right)\ket{\uparrow}+e^{i\beta}\sin\left(\frac{\alpha}{2}\right)\ket{\downarrow}$, where $\ket{\uparrow}$ and $\ket{\downarrow}$ are the eigenstates of $\sigma^{}_{z}$, a straightforward calculation gives the polarization along $\hat{\bold{m}}$ as (see more details in appendix \ref{Sec appA})
\begin{align}
\label{eq:29}&P^{}_{\hat{\bold{m}}}=P^{}_{\hat{\bold{n}}'}\left[\cos\alpha\cos\gamma+\sin\alpha\sin\gamma\cos\left(\beta-\delta\right)\right].
\end{align}
Equation \eqref{eq:29} reveals that $|P^{}_{\hat{\bold{m}}}|\leq |P^{}_{\hat{\bold{n}}'}|$ and the maximal polarization is obtained along the direction $\hat{\bold{n}}'$. Hence our model suggests that the spin polarization in the experiments, measured along the $z$-axis, is not the maximum polarization. One may achieve a larger polarization by measuring the spin of the transmitted electrons along a different direction $\hat{\bold{n}}'$. To characterize the direction $\hat{\bold{n}}'$ in more details, one should specify the spin rotation matrices $V^{}_{1}, \ldots, V^{}_{N}$ [see Eqs.\ \eqref{eq:2} and \eqref{eq:2b}]. In the next section we employ a helical geometry and introduce specific spin rotation matrices to study the spin polarization of helical molecules.
\section{Spin Polarization in a helical geometry} \label{Sec 2E}

The model presented in Sec.\ \ref{Sec 2A} allowed an arbitrary set of SOIs on the $N$ bonds within each unit cell along the helix (allowing e.g.\ different sites within the unit cell). The results presented so far depended only on the product matrix $\mathcal{V}$, and hence only on the angle $\theta$. In order to present an explicit relation between the SOI strength $\lambda$ and the angle $\theta$ we now present a simpler model, in which all the bonds within the unit cell are identical, except for the rotation around the helix axis.

For a helix of radius $R$ and pitch $h$ [see Fig.\ \ref{fig:Model}(a)] the vector $\mathbf{d}^{}_{n,n+1}$ is given by (the $z$-axis is chosen as the symmetry axis of the helix)
\begin{align}
\label{eq:3}&\mathbf{d}^{}_{n,n+1}=2R\sin\left(0.5\Delta\varphi\right)\left(-s^{}_{n}\hat{\mathbf{x}}+c^{}_{n}\hat{\mathbf{y}}\right)+\frac{h}{N}\hat{\mathbf{z}},
\end{align}
where $s^{}_{n}=\sin\left[\left(n+0.5\right)\Delta\varphi\right]$, $c^{}_{n}=\cos\left[\left(n+0.5\right)\Delta\varphi\right]$ and $\Delta\varphi=\pm 2\pi/N$ is the twist angle between nearest neighbors, with the plus (minus) sign corresponding to right-handed (left-handed) chirality. For a SOI induced by the confinement of the electron to the cylinder which contains the helix, the electric field can be assumed to be radial [see Fig.\ \ref{fig:Model}(a)],\cite{GAM12,Comment3}
\begin{align}
\label{eq:4}&\mathbf{E}^{}_{n,n+1}=E^{}_{0}\left(c^{}_{n}\hat{\mathbf{x}}+s^{}_{n}\hat{\mathbf{y}}\right).
\end{align}
Inserting Eqs.\ \eqref{eq:3} and \eqref{eq:4} into \eqref{eq:2b}, we obtain
\begin{align}
\label{eq:5}&\mathbf{K}^{}_{n}=\tilde{\lambda}\hat{\mathbf{e}}^{}_{n},
\end{align}
where $\tilde{\lambda}=\lambda E^{}_{0}\ell$ is a dimensionless parameter with $\ell=\sqrt{\left(h/N\right)^{2}+\left[2R\sin\left(0.5\Delta\varphi\right)\right]^{2}}$ being the length of each bond, and the unit vector $\hat{\mathbf{e}}^{}_{n}$ is
\begin{align}
\label{eq:6}&\hat{\mathbf{e}}^{}_{n}=\frac{1}{\ell}\left[\frac{h}{N}\left(-s^{}_{n}\hat{\mathbf{x}}+c^{}_{n}\hat{\mathbf{y}}\right)-2R\sin\left(0.5\Delta\varphi\right)\hat{\mathbf{z}}\right].
\end{align}

The spin rotation matrices are then related by a similarity transformation,
\begin{align}
\label{eq:30}&V^{}_{n}=U^{-n}V^{}_{N}U^{n},\nonumber\\
&U=e^{\pm i(\pi/N)\sigma^{}_{z}},
\end{align}
where the two signs correspond to right-handed and left-handed helix, respectively. Since $U^{N}=-\bold{1}$, we can express the unitary matrix $\mathcal{V}$ [Eqs.\ \eqref{eq:9} and
\eqref{eq:14}] as
\begin{align}
\label{eq:31}&\mathcal{V}\equiv V^{}_{N}V^{}_{N-1}\cdot\ldots\cdot V^{}_{1}=-(V^{}_{N}U)^{N}.
\end{align}
Using Eqs.\ \eqref{eq:2}, \eqref{eq:5}, \eqref{eq:6} and \eqref{eq:30}, we obtain
\begin{align}
\label{eq:32}&V^{}_{N}U=e^{i\phi\,\hat{\bold{n}}\cdot\boldsymbol\sigma}.
\end{align}
with (see appendix \ref{Sec appB} for more details)
\begin{align}
\label{eq:33}&\cos\phi=\cos\left(\frac{\pi}{N}\right)\cos\tilde{\lambda}+\frac{2R}{\ell}\sin^{2}\left(\frac{\pi}{N}\right)\sin\tilde{\lambda},\nonumber\\
&n^{}_{x}\sin\phi=\mp\frac{h}{N\ell}\sin\left(\frac{2\pi}{N}\right)\sin\tilde{\lambda},\nonumber\\
&n^{}_{y}\sin\phi=\frac{h}{N\ell}\cos\left(\frac{2\pi}{N}\right)\sin\tilde{\lambda},\nonumber\\
&n^{}_{z}\sin\phi=\pm\left[\sin\left(\frac{\pi}{N}\right)\cos\tilde{\lambda}-\frac{R}{\ell}\sin\left(\frac{2\pi}{N}\right)\sin\tilde{\lambda}\right].
\end{align}
Comparing Eqs.\ \eqref{eq:14}, \eqref{eq:31} and \eqref{eq:32} we identify $\theta=N\phi+\pi$, which gives the relation between $\theta$ and the SOI strength represented by $\tilde{\lambda}$.\cite{Comment4} We can then readily identify the direction of maximum polarization $\hat{\bold{n}}'=\left(\cos\beta\sin\alpha,\sin\beta\sin\alpha,\cos\alpha\right)$, defined by Eq.\ \eqref{eq:27} (see appendix \ref{Sec appB}). It should be emphasized that Eqs.\ \eqref{eq:33}, \eqref{eq:C3} and \eqref{eq:C4} depend only on the SOI strength and on the geometrical parameters characterizing the helix. They are independent of the complex site self-energy or the electron energy.

Let us discuss now the order of magnitude of $\theta$. In the limit $\tilde{\lambda}=\lambda E^{}_{0}\ell\ll 1$, one finds $\theta\approx 2N\sin\left(\pi/N\right)RE^{}_{0}\lambda$, so that $\theta$ is proportional to the SOI strength [see also discussion after Eq. (\ref{eq:14})]. In vacuum, one has $\lambda=e/\left(4m^{}_{e}c^{2}\right)$, where $e$ and $m^{}_{e}$ are the free electron charge and mass, respectively, and $c$ is the speed of light. If we estimate the radial confining field using typical values for the electric field in light atoms, $E^{}_{0}\approx 5\cdot 10^{11}\,$V/m (for which the SOI is of the order of a few meV), and set $R\approx 5\,$\AA, we obtain $\theta/\pi\approx 2.5\cdot 10^{-4}$. Similar values are expected for carbon nanotubes, since in this case the SOI is of the order of the SOI in atomic carbon (i.e.\ a few meV).\cite{KF08} However, the experimental results indicate that the observed spin polarization in chiral helical molecules cannot be explained using SOI of the order of a few meV.\cite{GB11,XZ11,MD13,EH15,NR12} The explanation of the observed spin polarization magnitude requires the SOI to be two or three orders of magnitude larger than the SOI in carbon systems. In the previous section we saw that our model yields significant values of the spin polarization (as measured e.g. in Refs. \onlinecite{GB11,XZ11,MD13,EH15}) assuming values of $\theta$, and therefore $\lambda$,  which are one or two orders of magnitude larger. Such an enhanced SOI may result from band structure effects (e.g.\ a smaller effective mass), as in narrow-gap semiconductors with strong Rashba SOI.\cite{Rashba,Winkler} However, the details of the mechanism for strong SOI in chiral helical molecules are yet to be clarified.

Equations \eqref{eq:C3} and \eqref{eq:C4} can be used to investigate the spin polarization along an arbitrary direction [see Eq.\ \eqref{eq:29}]. In particular, the experimentally
measurable spin polarization along the $z$-axis, $P^{}_{\hat{\bold{z}}}$, is of special interest. Experiments reveal that this quantity differs by a sign for molecules with opposite chiralities. This observation is analytically reproduced by our model; the polarization along the $z$-axis is given by Eq.\ \eqref{eq:29} with $\gamma=0$, that is
\begin{align}
\label{eq:36}&P^{}_{\hat{\bold{z}}}=P^{}_{\hat{\bold{n}}'}\cos\alpha.
\end{align}
Combining Eqs.\ \eqref{eq:33} and \eqref{eq:C3} one finds that $\cos\alpha$, and hence $P^{}_{\hat{\bold{z}}}$, reverses sign when the chirality is reversed.
\section{Discussion and Conclusions} \label{Sec 3}

We have presented a simple tight-binding model for electron transport in the presence of SOI in which spin polarization is generated by non-unitary hopping. Our model takes interference of electronic waves into account by considering hopping terms beyond NN ones. Although our model resembles that of Refs.\ \onlinecite{GAM12} and \onlinecite{GAM14}, there are several important differences. First, we consider only two hopping paths at each site. This simplifies the model and allows an analytical solution. One can then study in more detail the direction of maximum spin polarization. Second, we allow arbitrary SOI interactions on the bonds within each unit cell, which may imitate the internal structure of the helical molecules. This also makes our model applicable to other systems such as mesoscopic interferometers made of narrow-gap semiconductors with strong SOI. Third, our mechanism for the generation of a complex self-energy, due to electron leakage to side leads, differs from those of Refs.\ \onlinecite{GAM12} and \onlinecite{GAM14}.

At first look, our non-unitary effect may seem artificial. However, such a leakage out of the helical molecules and into the surrounding environment may in fact be the cause for the observed decay of the total current through the molecules with the molecule's length. Interestingly, our model also reproduces the experimentally observed increase in the spin polarization with the molecule's length. Our leakage may also mimic in a very simplified way complex effects such as electron capturing by DNA molecules,\cite{RSG05,MTZ13} as well as a variety of other phase-breaking processes. In general, a given microscopic phase-breaking mechanism (e.g., electron-electron or electron-phonon interactions) can be incorporated by a suitable choice of the complex self-energy function using the non-equilibrium Green's function formalism. However, phase-breaking processes are usually introduced by relatively
simple phenomenological models.\cite{MRG07} One phenomenological model for the inclusion of such processes was proposed by B\"{u}ttiker,\cite{Buttiker} in which phase-breaking processes are modelled by an additional electron reservoir coupled to the system via fictitious voltage probes, subject to the condition of zero net current. These B\"{u}ttiker probes remove electrons from the phase-coherent region and subsequently reinject them without any phase relationship. Another phenomenological model introduces phase-breaking processes by adding a spatially uniform pure imaginary potential to the Hamiltonian.\cite{EKB95} The latter model is similar to our approach. These two models were compared in Ref.\ \onlinecite{BPW97} and the limit in which they are equivalent was identified. In the context of the CISS effect, the authors of Refs.\ \onlinecite{GAM12} and \onlinecite{GAM14}
employed B\"{u}ttiker's method, whereas in Ref.\ \onlinecite{VS14} a phenomenological imaginary potential was introduced. In terms of the tight-binding formalism used here, both methods are expected to give a complex site self-energy which results in a spin splitting as discussed above.

The analytical solution of our simple model also suggests that one may achieve a larger CISS effect by measuring the spin polarization along an axis different from the $z$-axis,
conventionally used to measure the CISS effect. We identify the direction of this axis in terms of the SOI strength and the geometrical parameters characterizing the helix. Finally, we show analytically that the spin polarization along the $z$-axis changes sign when the chirality of the helix is reversed, also in agreement with experiments.

As we noted, coupling the molecule to two terminals and assuming time-reversal symmetry cannot yield a non-zero spin polarization. In some sense, our leakage model overcomes this problem by adding many more terminals, in which the electrons only go away from the molecule. An alternative way to achieve a non-zero spin polarization is to use a multi-terminal setup, namely a system with more than single input and output leads.\cite{FP06} In the context of our model one may consider $N$ input and output leads, connected to each of the $N$ sites in the leftmost and rightmost cells ($m=1$ and $m=M$). Such a model, which does not break unitarity, will be discussed elsewhere.

\begin{acknowledgments}
We acknowledge fruitful discussions with R. Naaman and K. Michaeli. This work was supported by the Israeli Science Foundation (ISF) and by the infrastructure program of Israel Ministry of Science and Technology under contract 3-11173.
\end{acknowledgments}
\appendix
\begin{widetext}
\section{Calculation of the polarization along an arbitrary axis} \label{Sec appA}
The eigenspinors of the spin projection
$\hat{\bold{n}}'\cdot\boldsymbol\sigma$ along the direction
$\hat{\bold{n}}'=\left(\cos\beta\sin\alpha,\sin\beta\sin\alpha,\cos\alpha\right)$
are
\begin{align}
\label{eq:appA1}&\ket{\hat{\bold{n}}'}=
\begin{pmatrix}
\cos\left(\frac{\alpha}{2}\right) \\
e^{i\beta}\sin\left(\frac{\alpha}{2}\right)
\end{pmatrix},\qquad
\ket{-\hat{\bold{n}}'}=
\begin{pmatrix}
\sin\left(\frac{\alpha}{2}\right) \\
-e^{i\beta}\cos\left(\frac{\alpha}{2}\right)
\end{pmatrix}.
\end{align}
Using the second of Eqs.\ \eqref{eq:25}, we obtain
\begin{align}
\label{eq:appA2}&\mathcal{T}\mathcal{T}^{\dag}=|t^{}_{\uparrow\uparrow}|^{2}\ket{\hat{\bold{n}}'}\bra{\hat{\bold{n}}'}+|t^{}_{\downarrow\downarrow}|^{2}\ket{-\hat{\bold{n}}'}\bra{-\hat{\bold{n}}'}=\nonumber\\
&\begin{pmatrix}
|t^{}_{\uparrow\uparrow}|^{2}\cos^{2}\left(\frac{\alpha}{2}\right)+|t^{}_{\downarrow\downarrow}|^{2}\sin^{2}\left(\frac{\alpha}{2}\right) & \frac{1}{2}e^{-i\beta}\sin\alpha\left(|t^{}_{\uparrow\uparrow}|^{2}-|t^{}_{\downarrow\downarrow}|^{2}\right) \\
\frac{1}{2}e^{i\beta}\sin\alpha\left(|t^{}_{\uparrow\uparrow}|^{2}-|t^{}_{\downarrow\downarrow}|^{2}\right)
&
|t^{}_{\uparrow\uparrow}|^{2}\sin^{2}\left(\frac{\alpha}{2}\right)+|t^{}_{\downarrow\downarrow}|^{2}\cos^{2}\left(\frac{\alpha}{2}\right)
\end{pmatrix}.
\end{align}
Thus, for an arbitrary direction
$\hat{\bold{m}}=\left(\cos\delta\sin\gamma,\sin\delta\sin\gamma,\cos\gamma\right)$
we find
\begin{align}
\label{eq:appA3}&\Tr\left[\mathcal{T}^{\dag}\mathcal{T}\right]=\Tr\left[\mathcal{T}\mathcal{T}^{\dag}\right]=|t^{}_{\uparrow\uparrow}|^{2}+|t^{}_{\downarrow\downarrow}|^{2},\nonumber\\
&\Tr\left[\mathcal{T}^{\dag}\left(\hat{\bold{m}}\cdot\boldsymbol\sigma\right)\mathcal{T}\right]=\Tr\left[\mathcal{T}\mathcal{T}^{\dag}\left(\hat{\bold{m}}\cdot\boldsymbol\sigma\right)\right]=\left[\cos\alpha\cos\gamma+\sin\alpha\sin\gamma\cos\left(\beta-\delta\right)\right]\left(|t^{}_{\uparrow\uparrow}|^{2}-|t^{}_{\downarrow\downarrow}|^{2}\right),
\end{align}
from which one obtains Eq.\ \eqref{eq:29}.
\section{Calculation of the maximum polarization direction $\hat{\bold{n}}'$} \label{Sec appB}
Using Eq.\ \eqref{eq:6} the matrix $V^{}_{N}$ is
\begin{align}
\label{eq:appB1}&V^{}_{N}=e^{i\tilde\lambda\hat{\mathbf{e}}^{}_{N}\cdot\boldsymbol\sigma}=\begin{pmatrix}
\cos\tilde\lambda\pm i\frac{2R}{\ell}\sin\left(\frac{\pi}{N}\right)\sin\tilde\lambda & \frac{h}{N\ell}e^{\mp\pi i/N}\sin\tilde\lambda \\
-\frac{h}{N\ell}e^{\pm\pi i/N}\sin\tilde\lambda &
\cos\tilde\lambda\pm
i\frac{2R}{\ell}\sin\left(\frac{\pi}{N}\right)\sin\tilde\lambda
\end{pmatrix}.
\end{align}
Combining the second of Eqs.\ \eqref{eq:30} and Eq.\
\eqref{eq:appB1}, we find
\begin{align}
\label{eq:appB2}&V^{}_{N}U=\left[\cos\left(\frac{\pi}{N}\right)\cos\tilde\lambda+\frac{2R}{\ell}\sin^{2}\left(\frac{\pi}{N}\right)\sin\tilde\lambda\right]\bold{1}\nonumber\\
&+i\begin{pmatrix}
\pm\left[\sin\left(\frac{\pi}{N}\right)\cos\tilde\lambda-\frac{R}{\ell}\sin\left(\frac{2\pi}{N}\right)\sin\tilde\lambda\right] & -i\frac{h}{N\ell}e^{\mp2\pi i/N}\sin\tilde\lambda \\
-i\frac{h}{N\ell}e^{\pm2\pi i/N}\sin\tilde\lambda &
\mp\left[\sin\left(\frac{\pi}{N}\right)\cos\tilde\lambda-\frac{R}{\ell}\sin\left(\frac{2\pi}{N}\right)\sin\tilde\lambda\right]
\end{pmatrix}.
\end{align}
Comparing this form with Eq.\ \eqref{eq:32} one arrives at Eqs.\
\eqref{eq:33}. Using Eq.\ \eqref{eq:27}, one can operate with
$V^{\dag}_{N}$ on $\ket{\pm\hat{\bold{n}}}$ to find the direction
of maximum polarization
$\hat{\bold{n}}'=\left(\cos\beta\sin\alpha,\sin\beta\sin\alpha,\cos\alpha\right)$.
A straightforward algebra gives
\begin{align}
\label{eq:C3}&\cos\alpha=\frac{h}{N\ell}\left[\frac{4R}{\ell}\sin^{2}\left(\frac{\pi}{N}\right)\sin^{2}\tilde\lambda-\cos\left(\frac{\pi}{N}\right)\sin(2\tilde\lambda)\right]n^{}_{x}\nonumber\\
&\mp\frac{h}{N\ell}\left[\frac{2R}{\ell}\sin\left(\frac{2\pi}{N}\right)\sin^{2}\tilde\lambda+\sin\left(\frac{\pi}{N}\right)\sin(2\tilde\lambda)\right]n^{}_{y}\nonumber\\
&+\left(1-\frac{2h^{2}}{N^{2}\ell^{2}}\sin^{2}\tilde\lambda\right)n^{}_{z},\nonumber\\
&\tan\beta=\frac{\pm An^{}_{x}+Bn^{}_{y}\mp
Cn^{}_{z}}{Dn^{}_{x}\pm En^{}_{y}+Fn^{}_{z}},
\end{align}
with
\begin{align}
\label{eq:C4}&A=\frac{h^{2}}{2N^{2}\ell^{2}}\cos\left(\frac{2\pi}{N}\right)\sin^{2}\tilde\lambda+\frac{R}{\ell}\sin\left(\frac{\pi}{N}\right)\sin(2\tilde\lambda),\nonumber\\
&B=\frac{1}{2}\left[\cos(2\tilde\lambda)+\frac{h^{2}}{N^{2}\ell^{2}}\left(1+\cos\left(\frac{2\pi}{N}\right)\right)\sin^{2}\tilde\lambda\right],\nonumber\\
&C=\frac{h}{N\ell}\left[\frac{R}{\ell}\sin\left(\frac{2\pi}{N}\right)\sin^{2}\tilde\lambda+\frac{1}{2}\sin\left(\frac{\pi}{N}\right)\sin(2\tilde\lambda)\right],\nonumber\\
&D=\frac{1}{2}\left[\cos(2\tilde\lambda)+\frac{h^{2}}{N^{2}\ell^{2}}\left(1-\cos\left(\frac{2\pi}{N}\right)\right)\sin^{2}\tilde\lambda\right],\nonumber\\
&E=\frac{h^{2}}{2N^{2}\ell^{2}}\sin\left(\frac{2\pi}{N}\right)\sin^{2}\tilde\lambda-\frac{R}{\ell}\sin\left(\frac{\pi}{N}\right)\sin(2\tilde\lambda),\nonumber\\
&F=\frac{h}{N\ell}\left[\frac{2R}{\ell}\sin^{2}\left(\frac{\pi}{N}\right)\sin^{2}\tilde\lambda+\frac{1}{2}\sin\left(\frac{\pi}{N}\right)\sin(2\tilde\lambda)\right].
\end{align}

\end{widetext}


\begin{references}
\bibitem{NWJM07} W. J. M. Naber, S. Faez, and W. G. van der Wiel, J. Phys. D: Appl. Phys. {\bf 40}, R205 (2007).
\bibitem{DVA} V. A. Dediu, L. E. Hueso, I. Bergenti, and C. Taliani, Nature Mater. {\bf 8}, 707 (2009); I. Bergenti, V. Dediu, M. Prezioso, and A. Riminucci, Philos. Trans. R. Soc. London, Ser. A {\bf 369}, 3054 (2011).
\bibitem{BC13} C. Boehme and J. M. Lupton, Nat. Nanotechnol. {\bf 8}, 612 (2013).
\bibitem{RSG06} S. G. Ray, S. S. Daube, G. Leitus, Z. Vager, and R. Naaman, Phys. Rev. Lett. {\bf 96}, 036101 (2006).
\bibitem{GB11} B. G\"{o}hler, V. Hamelbeck, T. Z. Markus, M. Kettner, G. F. Hanne, Z. Vager, R. Naaman, and H. Zacharias, Science {\bf 331}, 894 (2011).
\bibitem{XZ11} Z. Xie, T. Z. Markus, S. R. Cohen, Z. Vager, R. Gutierrez, and R. Naaman, Nano Lett. {\bf 11}, 4652 (2011).
\bibitem{MD13} D. Mishra, T. Z. Markus, R. Naaman, M. Kettner, B. G\"{o}hler, H. Zacharias, N. Friedman, M. Sheves, and C. Fontanesi, PNAS {\bf 110}, 14872 (2013).
\bibitem{EH15} H. Einati, D. Mishra, N. Friedman, M. Sheves, and R. Naaman, Nano Lett. {\bf 15}, 1052 (2015).
\bibitem{BDO13} O. Ben Dor, S. Yochelis, S. P. Mathew, R. Naaman, an Y. Paltiel, Nat. Commun. {\bf 4}, 2256 (2013).
\bibitem{GR12} R. Gutierrez, E. D\'{i}az, R. Naaman, and G. Cuniberti, Phys. Rev. B {\bf 85}, 081404(R) (2012).
\bibitem{GR13} R. Gutierrez, E. D\'{i}az, C. Gaul, T. Brumme, F. Dom\'{i}nguez-Adame, and G. Cuniberti, J. Phys. Chem. C {\bf 117}, 22276 (2013).
\bibitem{GAM12} A.-M. Guo and Q.-F. Sun, Phys. Rev. Lett. {\bf 108}, 218102 (2012).
\bibitem{GAM14} A.-M. Guo and Q.-F. Sun, PNAS {\bf 111}, 11658 (2014).
\bibitem{VS14} S. Varela, E. Medina, F. L\'{o}pez, and V. Mujica, J. Phys.: Condens. Matter {\bf 26}, 015008 (2014).
\bibitem{ME12} E. Medina, F. L\'{o}pez, M. A. Ratner, and V. Mujica, Europhys. Lett. {\bf 99}, 17006 (2012).
\bibitem{EAA13} A. A. Eremko and V. M. Loktev, Phys. Rev. B {\bf 88}, 165409 (2013).
\bibitem{GJ13} J. Gersten, K. Kaasbjerg, and A. Nitzan, J. Chem. Phys. {\bf 139}, 114111 (2013).
\bibitem{Dresselhaus} G. Dresselhaus, Phys. Rev. {\bf 100}, 580 (1955).
\bibitem{Rashba} E. I. Rashba, Fiz. Tverd. Tela (Leningrad) {\bf 2}, 1224 (1960) [Sov. Phys. Solid State {\bf 2}, 1109 (1960)]; Y. A. Bychkov and E. I. Rashba, J. Phys. C {\bf 17}, 6039 (1984).
\bibitem{Winkler} R. Winkler, {\it Spin-Orbit Coupling Effects in Two-Dimensional Electron and Hole Systems} (Springer-Verlag, Berlin, 2003).
\bibitem{MH92} H. Mathur and A. D. Stone, Phys. Rev. Lett. {\bf 68}, 2964 (1992).
\bibitem{MJS02} J. S. Meyer, V. I. Fal'ko, and B. L. Altshuler, in Nato Science Series II, Vol. 72, edited by I. V. Lerner, B. L. Altshuler, V. I. Fal'ko, and T. Giamarchi (Kluwer Academic Publishers, Dordrecht, 2002), p. 142.
\bibitem{MB04} B. Moln\'{a}r, F. M. Peeters, and P. Vasilopoulos, Phys. Rev. B {\bf 69}, 155335 (2004).
\bibitem{CR06} R. Citro, F. Romeo, and M. Marinaro, Phys. Rev. B {\bf 74}, 115329 (2006).
\bibitem{HN07} N. Hatano, R. Shirasaki, and H. Nakamura, Phys. Rev. A {\bf 75}, 032107 (2007).
\bibitem{AA11} A. Aharony, Y. Tokura, G. Z. Cohen, O. Entin-Wohlman, and S. Katsumoto, Phys. Rev. B {\bf 84}, 035323 (2011).
\bibitem{MS13b} S. Matityahu, A. Aharony, O. Entin-Wohlman, and S. Tarucha, New J. Phys.\ {\bf 15}, 125017 (2013).
\bibitem{BCWJ97} C. W. J. Beenakker, Rev. Mod. Phys. {\bf 69}, 731 (1997).
\bibitem{KAA05} A. A. Kiselev and K. W. Kim, Phys. Rev. B {\bf 71}, 153315 (2005).
\bibitem{BJH08} J. H. Bardarson, J. Phys. A {\bf 41}, 405203 (2008).
\bibitem{com} As shown in Ref.\ \onlinecite{BJH08}, this degeneracy is most easily proved by choosing a basis of scattering states in which the outgoing modes are the time reversed of the incoming modes. In this basis one finds that the scattering matrix $S$ is antisymmetric, $S^{T}=-S$, and the Kramers degeneracy of transmission eigenvalues follows from the resulting antisymmetry of the reflection amplitude matrix and the unitarity of the scattering matrix.
\bibitem{MS13} S. Matityahu, A. Aharony, O. Entin-Wohlman, and S. Katsumoto, Phys. Rev. B {\bf 87}, 205438 (2013).
\bibitem{AA02} A. Aharony, O. Entin-Wohlman, B. I. Halperin, and Y. Imry, Phys. Rev. B {\bf 66}, 115311 (2002).
\bibitem{Buttiker} M. B\"{u}ttiker, Phys. Rev. B {\bf 33}, 3020 (1986); M. B\"{u}ttiker, IBM J. Res. Dev. {\bf 32}, 72 (1988).
\bibitem{EKB95} K. B. Efetov, Phys. Rev. Lett. {\bf 74}, 2299 (1995).
\bibitem{RSG05} S. G. Ray, S. S. Daube, and R. Naaman, PNAS {\bf 102}, 14 (2005).
\bibitem{MTZ13} T. Z. Markus, A. R. de Leon, D. Reid, C. Achim, and R. Naaman, J. Phys. Chem. Lett. {\bf 4}, 3298 (2013).
\bibitem{OY92} Y. Oreg and O. Entin-Wohlman, Phys. Rev. B {\bf 46}, 2393 (1992).
\bibitem{Comment1} It should be noted that the addition of SOI on the bonds beyond NN ones allows for a finite spin polarization in a setup with more than a single input and output leads, even without fictitious side electrodes simulating phase-breaking processes or leakage. Such a model will be published elsewhere.
\bibitem{Comment2} For simplicity, we take the hopping amplitude on the bonds between the leads and the molecule to be $J^{}_{0}$. Other choices give qualitatively similar results.
\bibitem{Comment3} If the electron moves along a one-dimensional helical curve, the confining field has components both along the radial and the $z$-axis. However, since we assume the electron can hop along the $z$-axis (through the vertical bonds characterized by $\tilde{J}$), it is more appropriate to consider a radial confining field.
\bibitem{Comment4}  As stated after Eq.\ \eqref{eq:14}, in the limit $\lambda\rightarrow 0$ (and thus $\tilde{\lambda}\rightarrow 0$), one has $\mathcal{V}\rightarrow\bold{1}$. This is consistent with Eq.\ \eqref{eq:31}, since $U^{N}=-\bold{1}$; the minus sign is picked up by the spinors of fermions upon rotation by $2\pi$ in spin space.
\bibitem{KF08} F. Kuemmeth, S. Ilani, D. C. Ralph, and P. L. McEuen, Nature (London) {\bf 452}, 448 (2008).
\bibitem{NR12} R. Naaman and D. H. Waldeck, J. Phys. Chem. Lett. {\bf 3}, 2178 (2012).
\bibitem{MRG07} R. Golizadeh-Mojarad and S. Datta, Phys. Rev. B {\bf 75}, 081301(R) (2007).
\bibitem{BPW97} P. W. Brouwer and C. W. J. Beenakker, Phys. Rev. B {\bf 55}, 4695 (1997).
\bibitem{FP06} P. F\"{o}ldi, O. K\'{a}lm\'{a}n, M. G. Benedict, and F. M. Peeters, Phys. Rev. B {\bf 73}, 155325 (2006).

\end{references}
\end{document}